**Brightened emission of dark trions in transition-metal dichalcogenide monolayers**

V. Jindal[1*], K. Mourzidis[1*], A. Balocchi[1], C. Robert[1], P. Li[2], D. Van Tuan[2], L. Lombez[1], D. Lagarde[1], P. Renucci[1], T. Taniguchi[3], K. Watanabe[4], H. Dery[2,5†] and X. Marie[1,6†]

*[1]Université de Toulouse, INSA-CNRS-UPS, LPCNO, 135 Av. Rangueil, 31077 Toulouse, France*
*[2] Department of Electrical and Computer Engineering, University of Rochester, Rochester, New York 14627, USA*
*[3]International Center for Materials Nanoarchitectonics, National Institute for Materials Science, 1-1 Namiki, Tsukuba 305-00044, Japan*
*[4]Research Center for Functional Materials, National Institute for Materials Science, 1-1 Namiki, Tsukuba 305-00044, Japan*
*[5] Department of Physics and Astronomy, University of Rochester, Rochester, New York 14627, USA*
*[6] Institut Universitaire de France, 75231 Paris, France*

*The optical emission spectra of semiconducting transition-metal dichalcogenide monolayers highlight fascinating recombination processes of charged excitons (trions). When charge tunable* $WSe_2$ *monolayers are moderately doped with electrons, a strong luminescence peak emerges just below the well-understood spectral lines associated with the recombination of negatively charged bright and dark trions. Despite previous investigations, its origin remains elusive. Here, we demonstrate that this luminescence peak is the result of electron-electron assisted recombination that brightens the dark trion emission. Supporting evidence for this second-order recombination process comes from identifying the brightened emission of positively charged dark trions when the monolayer is electrostatically doped with holes. Remarkably, the discovered hole-hole assisted luminescence peak emerges in the near infrared, about 500 meV below the well-studied spectral region of excitons and trions. In magneto-photoluminescence experiments we find that the g-factor of this new transition (g = +4) has an opposite sign compared to the well-known g-factor of neutral or charged excitons. This allows us to propose a mechanism of brightening of the positively charged dark trion involving the* $\Gamma$ *valence band.*

*These authors contributed equally to this work.

† hdery@UR.Rochester.edu , marie@insa-toulouse.fr

## I. INTRODUCTION

The optical properties of atomically thin semiconductors based on transition-metal dichalcogenides (TMDs) are dominated by robust exciton complexes which give rise to rich physics associated with various interactions and correlation effects [1–3]. Charge-tunable $WSe_2$ monolayers are possibly the most studied model system due to their relatively high quality. When electrostatically doped with electrons, the luminescence spectra of the monolayer exhibit well identified peaks corresponding to the optical recombination of bright and dark negatively charged excitons (trions) [4–7] . Surprisingly, between slight and moderate densities of resident electrons, the most intense luminescence line does not correspond to these direct optical recombination channels: an intense luminescence component lying ~10 meV below the dark trion has been clearly identified by many groups in charge tunable $WSe_2$ monolayer devices [4,8–11] . Two fundamental questions arise: which exciton complex is associated with this optical recombination? And what is the nature of the recombination process? Several different interpretations were proposed. It was suggested that the line could result from the trion's fine structure [4], the recombination of double charged trion $X^{2-}$, [9] the coupling between excitons and short-range intervalley plasmons [12] or the consequence of the interaction between dark and bright trions [13,14].

Here, we elucidate the origin of this luminescence line and demonstrate that its relatively intense emission is the result of a second-order transition that involves dark trions. A resonant effect associated with the electron-electron interaction yields radiative recombination along with spin-conserving intervalley transition of the left behind electron to the upper conduction band (the electron that originally resides in the same valley of the hole). As a result of energy conservation and translation symmetry, the emitted photon is shifted to lower energy compared to the dark trion by a value equal to the conduction band (CB) spin-orbit splitting $\Delta_c$. We find $\Delta_c$= 12.0 $\pm$ 0.5 meV. The interpretation is further strengthened by the fact that we observe and interpret the same type of recombination channel for the positively charged trion when the monolayer is electrostatically doped with holes. It results in a near infrared emission line, as the emission is now down-shifted by the splitting in the valence band, which is much larger than the CB one. The measurement of the $g$-factor of this new transition ($g$ =+ 4), *i.e.* opposite sign to the $g$-factor of the bright neutral and charged exciton, shows that the brightening process of the positively charged dark trion cannot be explained by a simple hole-hole assisted recombination involving the A and B valence band states in K valley. In contrast, the $g$-factor value can be well interpreted if the top of the valence band in Γ valley is considered. We propose a mechanism involving the interplay between the B valence band in K valley and the top of VB in Γ valley, which have nearly the same energy as demonstrated in recent Angle Resolved Photoemission Spectroscopy (ARPES) measurements [15,16].

## II. EXPERIMENTAL RESULTS

We investigated high-quality charge tunable $WSe_2$ devices as sketched in Fig. 1(a). [17,18].

*Sample fabrication*. The van der Waals heterostructure is made of an exfoliated ML-$WSe_2$ embedded in high quality hBN crystals using a dry stamping technique in the inert atmosphere of a glove box [19,20] . The layers are deterministically transferred on top of a $SiO_2$/Si substrate with Ti/Au electrodes patterned by photolithography. Flux-grown $WSe_2$ bulk crystals are purchased from 2D semiconductors. We use few layers of graphene exfoliated from a Highly Ordered Pyrolytic Graphite (HOPG) bulk crystal for the back gate and to contact the ML-$WSe_2$.

*Experimental set-up***.** Low temperature photoluminescence measurements were performed in a home built micro-spectroscopy set-up built around a closed-cycle, low vibration cryostat (T=4 K). The sample is excited by a He-Ne laser (632.8 nm). The PL signal is dispersed by a spectrometer and detected by a silicon based charged-coupled device camera [21]. Magneto- PL experiments in magnetic fields up to 6.5 T were also carried out in a magneto-cryostat at T=4 K. The sample is excited by a He-Ne laser (632.8 nm) with linear polarization and both circular σ+ and σ- polarized PL signals are detected using a λ/4 plate and an analyzer in the detection path. Unless otherwise stated, the excitation power is typically ~50 μW, focused to a spot size of ~1 μm diameter.

A voltage bias applied between the monolayer and the back gate allows us to electrostatically dope the monolayer. The doping densities in either the electron- or hole-doped regimes are such that a change of 1 Volt in voltage bias corresponds to a typical change of $0.8 \times 10^{11}$ $cm^{-2}$ in charge density [22]. We deliberately remain in the low-density regime to avoid collective phenomena, renormalization effects, and the emergence of hexcitons [23]. In this regime, the three-particle trion picture is an adequate description [24,25]. Continuous-wave polarization-dependent micro-photoluminescence (PL) experiments are performed in a closed cycle cryostat (T = 4 K).

### A. Brightened emission of the negatively charged dark trion.

Figure 1(b) displays the PL color plot as a function of bias voltage and photon energy. In the charge-neutral regime (V ~ - 0.5 volts), indicated by the horizontal dashed line, the recombination of bright ($X^0$) and dark ($D^0$) excitons dominate the PL spectrum [26,27]. When electrons are added to the monolayer (positive voltages), the PL intensity of the neutral exciton vanishes and we observe the well-known recombination of intravalley (singlet) $X_S^-$ and intervalley (triplet) $X_T^-$ negatively charged trions, composed of two electrons and one hole [7,28] . As shown in Fig. 2(a), the triplet trion is composed of a photogenerated electron-hole pair, made of an electron in the top CB and a missing electron in the topmost VB of the same $K^+$ valley, and the pair is bound to a resident electron in the bottom CB of the time-reversed valley at $K^-$. For the singlet trion $X_S^-$, the three particles reside in the same valley, as shown in Fig. 2(b). The emission of the dark trion ($D^-$) is also detected at lower energy. As sketched in Fig. 2(c), $D^-$ is composed of two electrons in the lowest conduction bands at K+ and K-, and a hole at the top of the valence band. Its radiative recombination for in-plane polarized light is forbidden. However it is allowed for z-polarized light [10,11] ; the z-axis is perpendicular to the monolayer plane. The clear observation of $D^-$ in the PL spectrum, as shown in Fig.1, is the consequence of using high numerical aperture objective (NA=0.82) which allows us to detect the z-polarized luminescence component [26]. Importantly, we observe a bright line, labelled as $D_B^-$, lying 12 meV below $D^-$. This intense line was observed by many groups, but its origin remained unclear [4,8,10,11]. It was shown that this line is co-polarized with the laser, following circularly-polarized laser excitation and that its $g$-factor is about -4.5, *i.e.* close to the $g$-factor of bright trions or excitons [27]. We have checked that its intensity varies linearly with the excitation power in the range 1-50 μW. We interpret this line as the brightened emission of the dark trion $D^-$, resulting from exciton-electron interactions. The corresponding second-order recombination mechanism is summarized in Fig.2(d). The light is generated from recombination of the indirect exciton component composed of a conduction-band electron in $K^-$ and a hole in $K^+$ [29] To respect translation symmetry, the second electron of the trion ends-up in the opposite top valley (*i.e.*, Coulomb induced intervalley spin-conserving transition). As a result of energy conservation, the light emission is shifted to lower energy compared to the $D^-$ transition by an amount equal to the conduction band spin-obit splitting $\Delta_c$. The measured energy difference between $D^-$ and $D_B^-$ in Fig. 1(a) yields a direct measurement of the conduction band spin orbit splitting; we find $\Delta_c$ ~ 12 meV, in agreement with recent determinations [22,30]. As the $D_B^-$ transition involves electronic transitions to the upper conduction band states, this explains that the associated $g$-factor is that of the bright trion (and not of the dark excitonic species) [27] . The schematics in Fig. 2(d) also illustrate why the emission of $D_B^-$ is co-polarized with the laser following circularly polarized excitation, as observed experimentally [14]. We emphasize that contrary to the direct recombination process of bright or dark trions, the non-recombining electron in the dark trion is not merely a spectator during the $D_B^-$ emission. Though this $D_B^-$ emission results from a second-order process, its intensity is rather significative thanks to the resonance effect involving states in two nearby conduction bands. The interpretation of the $D_B^-$ transition is strongly reinforced by the observation of an equivalent emission, linked to the same type of Coulomb induced second-order mechanism of positively charged trions.

### B. Brightened emission of the positively charged dark trion.

Figure 1(b) shows that in the hole-doped regime (negative voltages), the spectrum is dominated by the recombination of the bright positive trion $X^+$ at 1.688 eV and the dark one $D^+$at 1.654 eV, as already observed by many groups [8,9,31]. Remarkably, when holes are injected in the monolayer, a new line shows up in the near-infrared region of the spectrum - beyond 1 μm wavelength. Figure 3(a) displays the PL spectrum in this region, which clearly evidences a new line, labelled $D_B^+$ at 1.175 eV; this line is absent in the neutral regime and its intensity rises with the voltage increase. We also note in the inset of Figure 3(a) the presence of 3 additional emission line at lower energy compared to $D_B^+$. The energies of these lines (1.138 eV, 1.120 eV and 1.083 eV) correspond respectively to the well-known zero-phonon optical recombination, TA and TO phonon-assisted optical recombination between free conduction band electrons and holes in the acceptor impurity band of the p-doped Silicon substrate [32] . As expected, the intensity of these silicon lines does not change with the applied voltage. We emphasize that no other lines show up with the applied voltage in the detected energy range (inset of Fig.3(b)). Figure 3(b) shows that the $D_B^+$ intensity is commensurate with that of the dark trion $D^+$. This peak was not detected before, probably because its emission lies at much lower energies compared to those associated with other exciton species. The inset of Fig. 3(b) shows that the intensity of $D_B^+$ rises with the excitation laser power, showing no sign of saturation, ruling out a possible interpretation based on impurity recombination.

The energy difference between the resonances $D_B^+$ and $D^+$ is 478 meV (Fig. 3(c)). Notably, it is very close to the valence band spin-orbit splitting which has been calculated [33] and measured in ARPES ( $485 \pm 10$ meV) [15] . This is a strong argument to interpret the emission of $D_B^+$ as the brightened emission of the dark trion $D^+$, following a second-order recombination mechanism like the one described above for the negatively charged dark trion $D^-$. However, as we will show below, the mechanism of brightening is a little bit more involved.

In order to get more information about this new transition, we have measured the *g*-factor in a longitudinal magnetic field. Right (σ+) and left (σ−) circularly polarized luminescence spectra of $D_B^+$ have been measured for a magnetic field varying between $B$ = −6.5 and +6.5 T, applied perpendicularly to the monolayer plane. The excitation laser is linearly polarized, and the gate voltage is - 5 V. For comparison, we have also recorded the magneto-PL spectra of the positively charged ($X^+$) exciton in the same magnetic field range. Fig. 4(a) and 4(b) display the $X^+$ and $D_B^+$ PL spectra for a magnetic field B= + 6 T. The Zeeman splitting $\Delta E = E_{\sigma+} - E_{\sigma-}$ is extracted by fitting the emission lines with Lorentzian functions. We plot in Fig. 4(c) the full magnetic field dependence of ΔE for both $X^+$ and $D_B^+$ transitions. This yields the determination of the $g$-factor as $\Delta E = g\mu_B B$, where $\mu_B$ is the Bohr magneton.

For the new $D_B^+$ transition, we get $g_{D_B^+} = +4.0 \pm 0.4$. Remarkably it has an opposite sign compared to the $g$-factor of the positively charged trion $g_{X^+} = -3.3 \pm 0.2$ (see the reversed position of σ+ and σ− spectra in Fig. 4(a) and 4(b), as well as the different slopes in Fig. 4(c)).

## III. DISCUSSION

Let us consider first that the emission from $D_B^+$ results simply from the second-order transition described above, a recombination similar to the brightening of $D^-$ but involving the two K valley valence bands split by the spin-orbit splitting $\Delta_v$. The process is sketched in Fig. 5, and it starts with a dark trion $D^+$ comprising two holes in the $K^+$ and $K^-$ top valleys of the valence band and one electron at the bottom of the conduction band (Fig. 5(a)). Owing to the hole-hole interaction, the $D_B^+$ near-infrared recombination is facilitated by radiative recombination of the momentum-indirect electron-hole component of the trion. Energy and wavevector conservation rules are fulfilled thanks to the fact that the second hole ends up in the lower-energy valence band (see the two anti-parallel arrows in Fig. 5(b)). As a consequence, the energy difference between the $D^+$ and $D_B^+$ lines with this mechanism would correspond to the valence band spin-orbit splitting $\Delta_v$.

We can now check if this scenario is consistent with the measured value of the g-factor of the $D_B^+$ transition. The g factors of the bottom and top CB in $K^+$ valley are $g_{c_1}$ and $g_{c_2}$ respectively, whereas the $g$-factors of the top and bottom VB in the same $K^+$ valley are denoted $g_{v_1}$ and $g_{v_2}$ (Fig. 5(b)). We recall that for a given CB or VB, $g(K^-)$=- $g(K^+)$ [34–36]. In this first scenario, the $D_B^+$transition energy is the difference between the energy of the indirect optical recombination of the electron in the bottom CB and the hole in the opposite valley VB (denoted $E_{indirect}$) and the energy $\Delta_v$, corresponding to the spin-conserving transition from the top VB in $K^+$ to bottom VB in $K^-$, or vice and versa (denoted $E_{inter-VB}$) :

$$E^{D_B^+} = E_{indirect} - E_{inter-VB} \quad (1)$$

In the presence of an external longitudinal magnetic field B, each band will experience an energy shift $\delta E = g\mu_B B$, with opposite sign in $K^+$ and $K^-$ valley. The energy variation for the σ+ transition (corresponding to the configuration of Fig. 5(b) ) writes [17] :

$$\delta E^+ = \delta E^+_{indirect} - \delta E^+_{inter-VB} \text{ with } \delta E^+_{indirect} = (g_{c_1} + g_{v_1})\,\mu_B B \text{ and } E^+_{inter-VB}\,(g_{v_1} + g_{v_2})\,\mu_B B,$$

yielding simply:

$$\delta E^+ = (g_{c_1} - g_{v_2})\,\mu_B B \quad (2)$$

Considering the other configuration with the electron lying in the $K^-$ valley, one gets the energy variation with magnetic field for the σ– transition:

$$\delta E^- = (-g_{c_1} + g_{v_2})\,\mu_B B \quad (3)$$

By subtracting equation (2) and (3), we obtain the Zeeman splitting for the $D_B^+$transition in this scenario:

$$\Delta E^{D_B^+} = 2(g_{c_1} - g_{v_2})\,\mu_B B \quad (4)$$

One can easily verify that the same reasoning for the $D_B^-$transition with the electron-electron assisted recombination summarized in Fig. 2(d) yields:

$$\Delta E^{D_B^-} = 2(g_{c_2} - g_{v_1})\mu_B B \quad (5)$$

The g-factor of the bottom (top) CB and VB were determined experimentally [17,37] (values in agreement with calculations [38–41] ) : $g_{c_1}$=0.86, $g_{c_2}$=3.84 and $g_{v_1}$=6.1, $g_{v_2}$=2.8.

Using equation (4), this yields a g factor of the $D_B^+$ transition of $g = 2(g_{c_1} - g_{v_2})$ ~ - 3.9. With an opposite sign, it is strikingly in disagreement with our measured value $g_{D_B^+}$ = +4.0 $\pm$ 0.4. In contrast, equation (5) for $D_B^-$ gives a calculated value $g$ ~ -4.5, in excellent agreement with the measured one $g_{D_B^-} = -4.5$ [27]. This shows that the brightening process of the positively charged trion does not result simply from a second-order process involving the K valleys in the valence band and there are some differences with the brightening process of the negatively charged exciton.

An important experimental difference between the emission lines $D_B^-$ and $D_B^+$ is that the linewidth of $D_B^+$ is about ten times larger than that of $D_B^-$ (Fig. 3(c)). The measured full width at half maximum of $D_B^+$ is ~24 meV, showing no dependence on the hole density in the studied range between 0 and ~ $5.10^{11}\ cm^{-2}$ . We attribute the large broadening to the ultrafast lifetime of the free hole in the bottom valence band. The ultrafast lifetime could stem from efficient intervalley hole scattering between the nearly degenerate valence band valley at Γ and the B valleys at ±K. (see Fig. 5(b)), as demonstrated in ARPES measurements and band structure calculations [15,16,42]. Such efficient scattering has no parallel in the conduction band. A spin-conserving intervalley transition mediated by $K_1$ phonon can lead to ultrafast transition of the hole from the bottom B valley to the Γ valley, whereas equivalent transition with $K_3$ mode in the conduction band is hampered by phonon bottleneck ($\Delta_c > E_{K3}$). We also emphasize that the Γ valley VB is characterized by a much larger density of states compared to the K valley as a consequence of heavier mass [33].

So, we propose that the brightening process of the positively charged dark trion $D^+$ results from a hole-hole in K valley to the VB in Γ valley (Fig. 5(b) and 5(c)), and not a spin-conserving transition from the top VB in $K^+$ to the bottom VB in $K^-$ as we described above.

A key argument to support this interpretation is the g factor of the transition, considering the involvement of the Γ valley. Following the same procedure as the one described by equations (1-3) and considering now that the transition for the left-behind hole occurs between the top VB in K valley to the VB in Γ valley, one can get the expression of the $D_B^+$ Zeeman splitting:

$$E = 2(g_{c_1} + g_{\Gamma})\mu_B B) \qquad (6)$$

, *where* $g_\Gamma$ is the $g$-factor of the Γ valley.

It turns out that $g_\Gamma$ can be deduced from the detailed calculations performed by Förste et al [40]. These authors determined the $g$-factor of the well-known transitions associated to the recombination of bright and dark neutral excitons and indirect ($K^+K^-$) neutral excitons, in very good agreement with the measured values. With the same approach, they also predicted the $g$-factor of the indirect transitions with an electron in $K^+$ or $K^-$ and a hole in the Γ valley. From these calculations, we can deduce $g_\Gamma \sim 1$. Using equation (6) this yields a calculated value $g = 2(g_{c_1} + g_\Gamma) \sim +3.7$. Remarkably, both the amplitude and the sign are in agreement with the measured $g$-factor of the $D_B^+$ transition ($g_{D_B^+} = +4.0 \pm 0.4$, Fig. 4(c)), a strong support for the brightening process associated to the hole-hole assisted recombination involving the Γ valley described in Fig. 5. Following this interpretation, the measured energy difference between the $D_B^+$ and $D^+$ transition, (478 meV) see Fig. 3(c), corresponds to the energy difference between the top K valley valence band and the Γ valley.

The interpretation of the $D_B^+$ line is further confirmed by measurements of the polarized luminescence following circularly polarized excitation.

The circularly polarized luminescence of $D_B^+$ is measured following circularly polarized laser excitation with helicity $\sigma^+$ (B=0). Figure 6(a) shows that the PL intensity counter-polarized with respect to the laser has a stronger intensity than the co-polarized PL component ($I^{\sigma+} < I^{\sigma-}$). The resulted PL polarization degree is $P_c = \frac{I^{\sigma+} - I^{\sigma-}}{I^{\sigma+} + I^{\sigma-}} \approx -0.25$. This negative PL polarization can be well explained by the scenario summarized in Fig. 6(b). The $\sigma^+$ laser excitation yields the initial photogeneration of a bright positively charge trion $X^+$ composed of an electron in the $K^+$ valley of the upper conduction band and two holes in $K^+$ and $K^-$ valley respectively. The energy relaxation of the electron to the lowest CB leads to the formation of the dark trion $D^+$. This can occur through a spin conserving process implying inter-valley $K_3$ phonon or an intra-valley relaxation based on $\Gamma_5$ zone-center phonon [27,43]. As was shown in refs. [18,27] , the inter-valley process is dominant, meaning that the largest population of dark trions $D^+$ following $\sigma^+$ laser excitation has electrons lying in the $K^-$ valley. Following hole-hole interaction, the radiative recombination will occur in the $K^-$ valley, as shown in Fig. 6(b). Consequently, a negative PL circular polarization of the $D_B^+$ line is expected, as we observe experimentally.

In summary, we have identified new recombination channels for dark trions in $WSe_2$ monolayers. The dependence of the transitions as a function of the doping, excitation power or external magnetic field allow us to propose a brightening process for both negatively and positively charged dark excitons based on a charge-charge assisted recombination process. Our interpretation explains well the energy and g-factor of the transitions; for the brightened emission of the positively charged dark trion, the process involves the interplay between the K and Γ valence band valleys. We believe that this kind of brightening mechanism of dark exciton species evidenced here in a $WSe_2$ monolayer could also occur in TMD homo- and hetero-structures.

**Acknowledgements**: This work was supported by the Agence Nationale de la Recherche under the program ESR/EquipEx+ (Grant No. ANR-21-ESRE- 0025) and ANR projects ATOEMS, IXTASE and PEPR SPIN. Work at the University of Rochester was supported by the DOE Basic Energy Sciences, Division of Materials Sciences and Engineering, under Award No. DE-SC0014349.

**Figures**

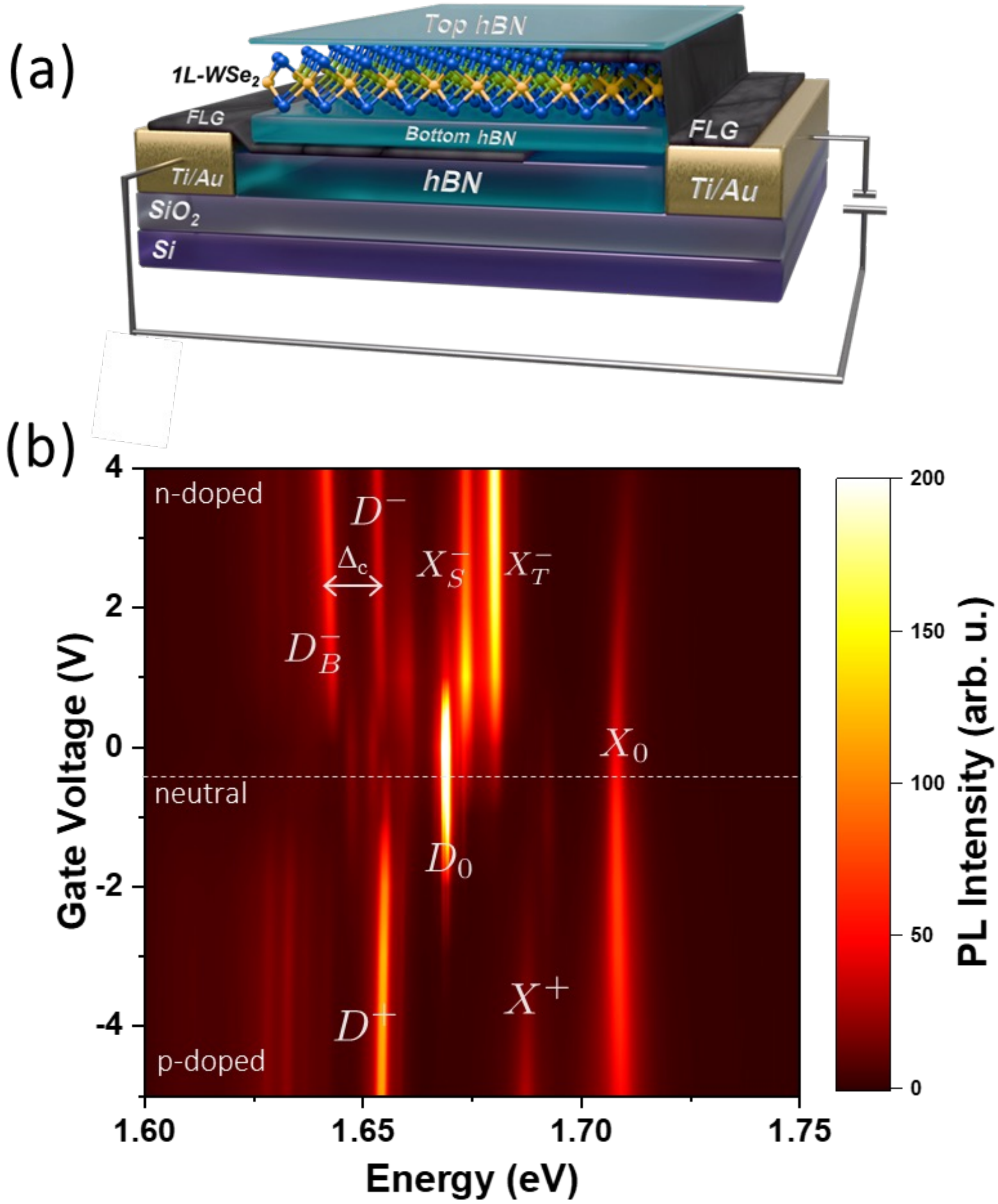

**Fig.1 Photoluminescence as a function of the gate voltage.** **(a)** Sketch of the charge adjustable $WSe_2$ monolayer device. (**b)** Photoluminescence intensity as a function of the gate voltage displaying the recombination of the different exciton complexes; the excitation energy is 1.96 eV. The brightened negatively charged exciton resulting from a second-order recombination process is labelled $D_B^-$ .

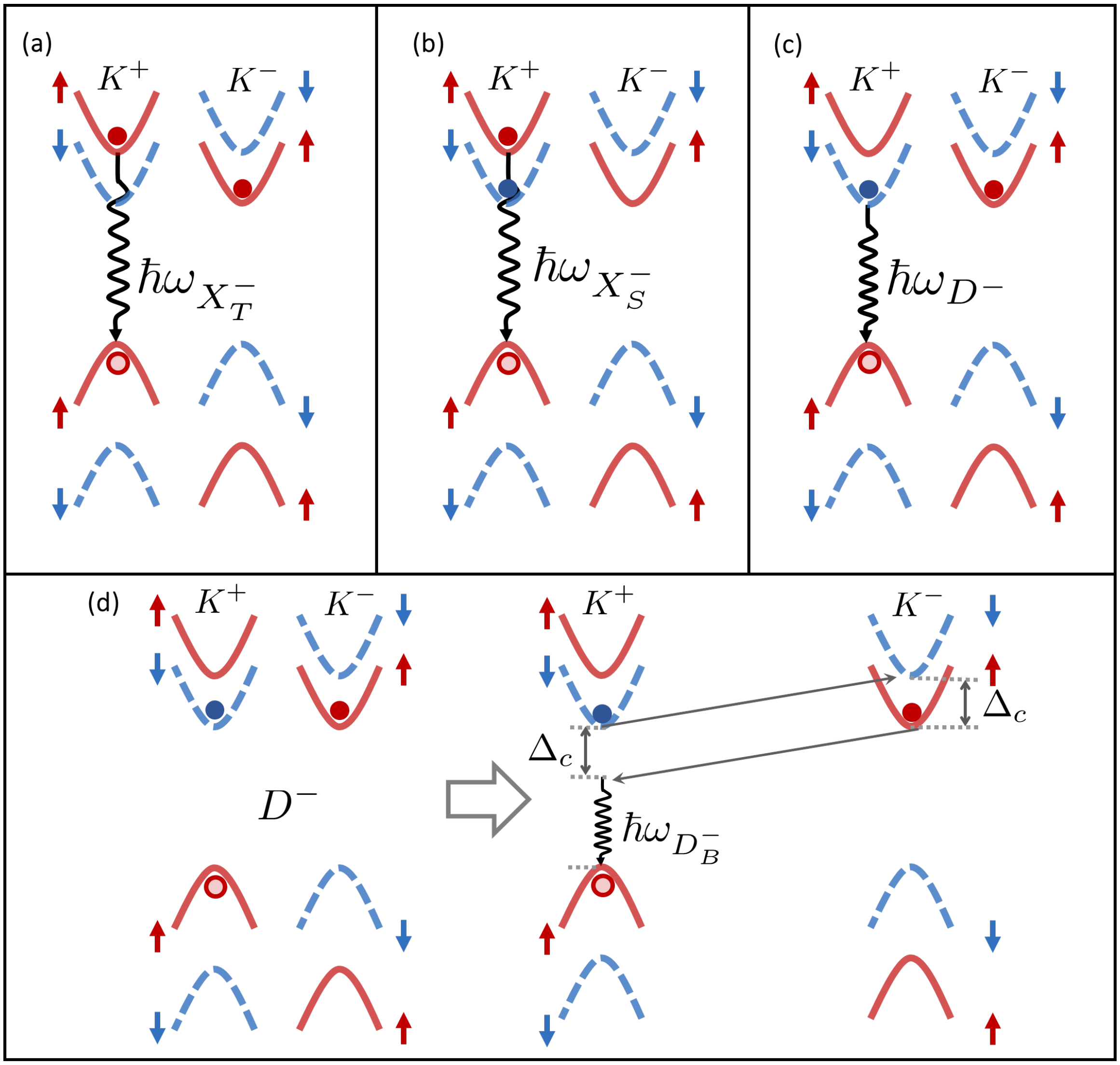

**Fig. 2 Negatively charged exciton recombination channels.** **(a)** and **(b)** Schematic of the bright triplet ($X_T^-$) and singlet ($X_S^-$) trion emission. **c** sketch of the dark ($D^-$) trion radiative recombination which is z-polarized. **d** Illustration of the second order transition starting from the dark trion ($D^-$) and yielding the radiative emission of the brightened dark trion ($D_B^-$) line: the light is generated from recombination of the indirect exciton component composed of the conduction band electron in $K^-$ and hole in $K^+$ valley. As a result of energy and wavevector conservation, the second electron of the trion ends-up in the opposite top valley (Coulomb induced inter-valley spin-conserving transition) and the process yields a light emission shifted to lower energy compared to the $D^-$ transition by an amount equal to the conduction band spin-obit splitting $\Delta_c$. In contrast to the direct optical recombination of bright or dark trion (as in **(a)**, **(b)** or **(c)**), the second electron here is no more a spectator.

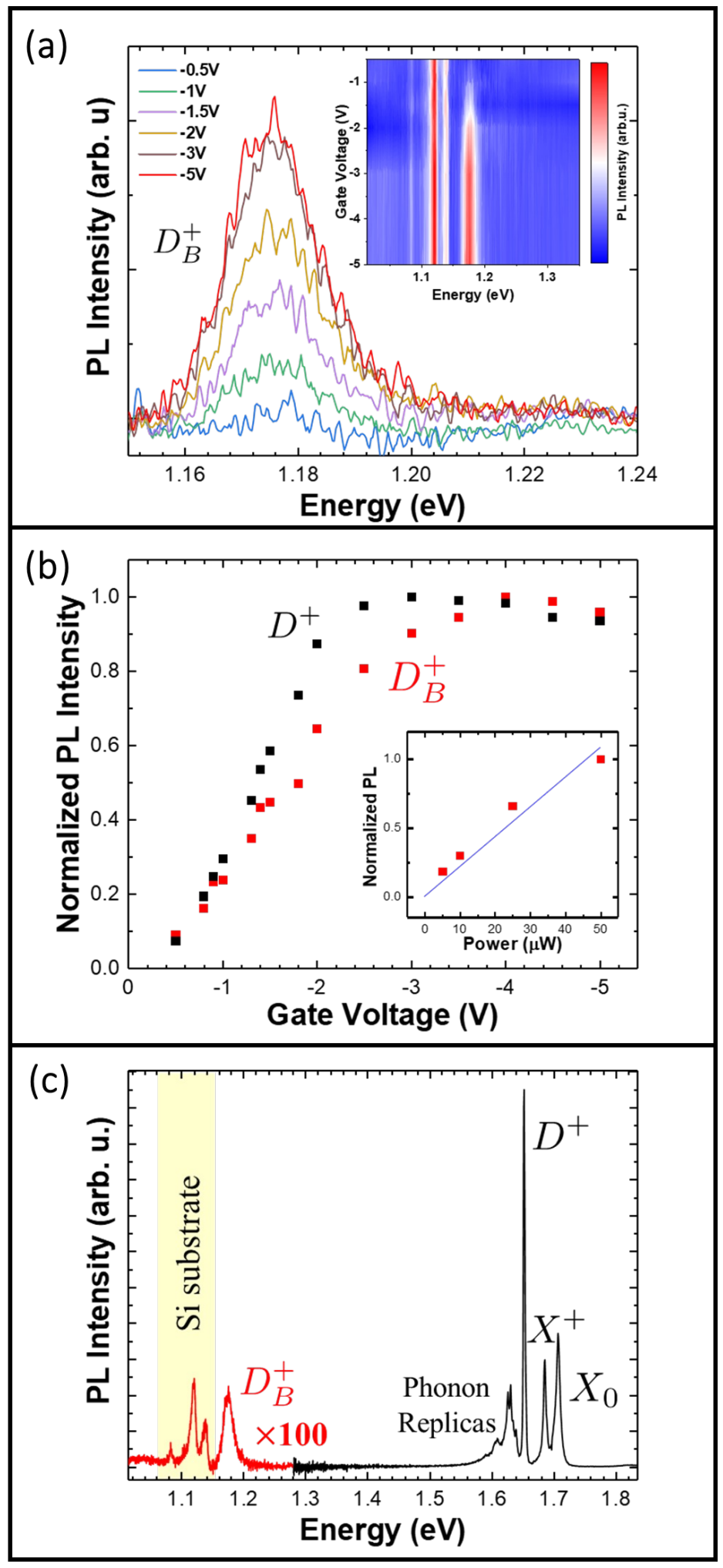

**Fig. 3 Brightened emission of the positively charged dark trion.** **(a)** PL intensity of the $D_B^+$ line as a function of the gate voltage; note the energy range which corresponds to emission wavelengths beyond 1 μm; inset: color plot of the photoluminescence spectra as a function of the gate voltage displaying the three well-identified lines from the silicon substrate (see text) in addition to the positively charged dark trion $D_B^+$ . **(b)** Normalized PL intensity of the dark $D^+$ and brightened dark $D_B^+$ trions as a function of the gate voltage; the inset displays the excitation power dependence of the $D_B^+$ line for a voltage V=-5 volt. **(c)** Photoluminescence spectrum displayed on a large energy range showing the energy difference between $D^+$ and $D_B^+$.

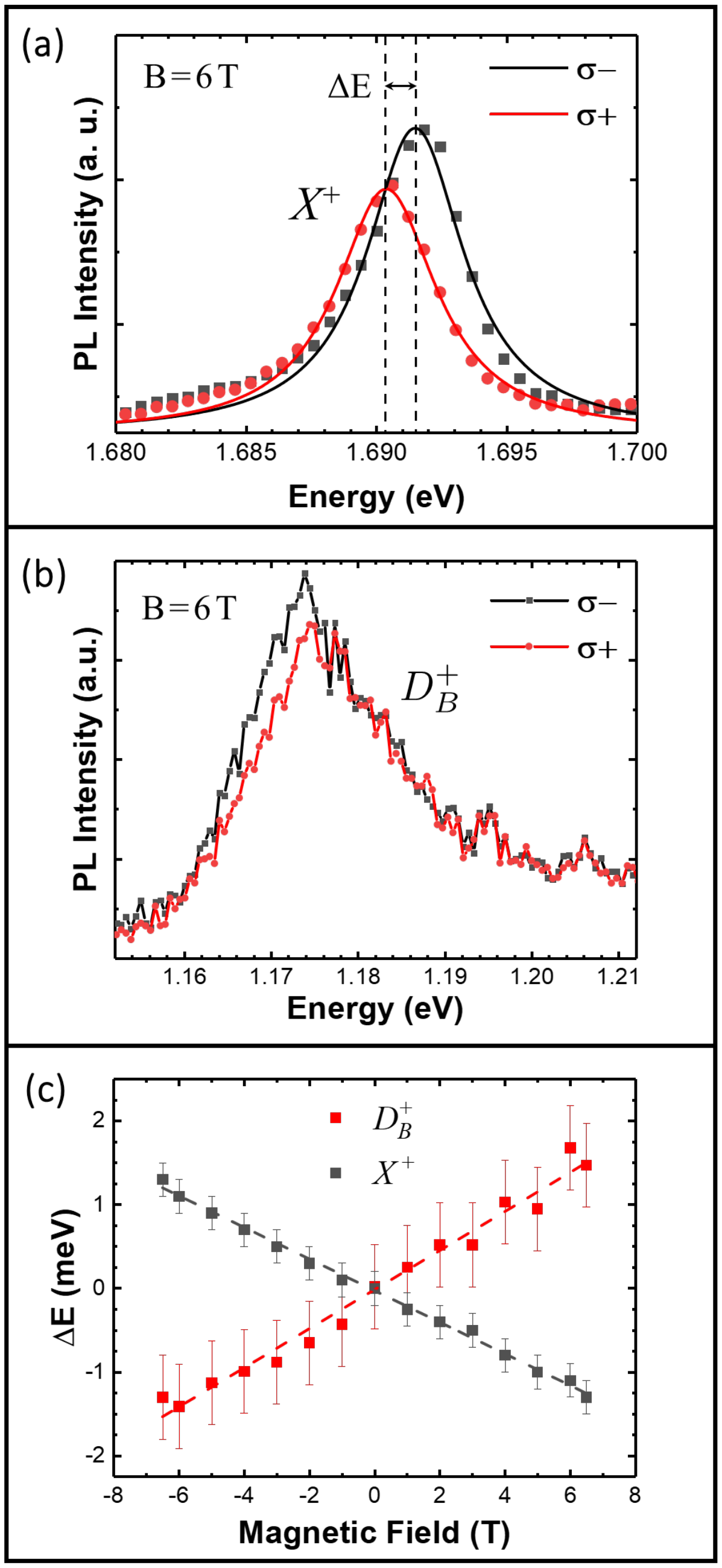

**Fig. 4 Magneto-photoluminescence measurements.** PL spectra at B= 6T for σ+ (red) and σ− (black) polarized detection for **(a)** the $X^+$ and **(b)** $D_B^+$ transition. **(c)** Energy splitting between the σ+ and σ− polarized PL components for both $X^+$ and $D_B^+$ lines as a function of magnetic field.

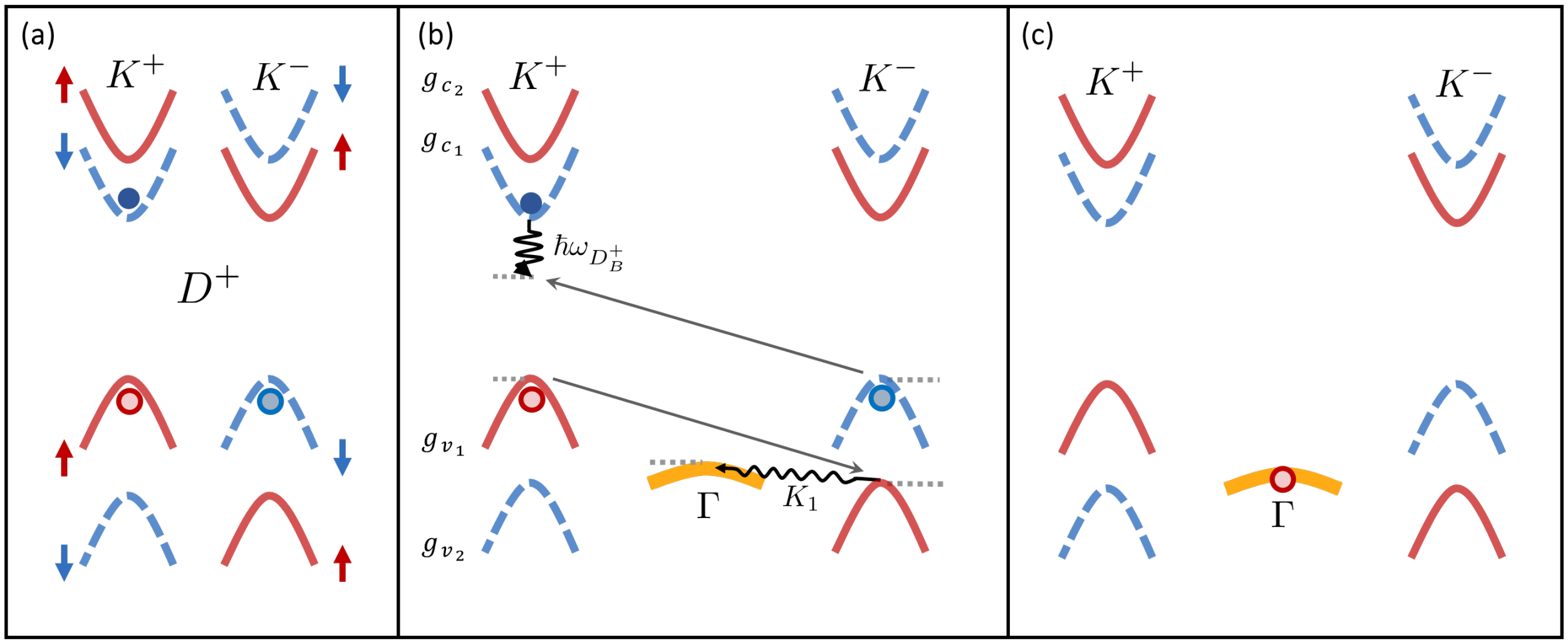

**Fig. 5 Brightening mechanism of the positively charged dark trion.**
(**a**) The dark trion $D^+$ is composed of 2 holes lying in $K^+$ and $K^-$ valleys at the top of the valence band (A) and one electron lying at the bottom of the conduction band. **(b)** The $D_B^+$ near-infrared recombination results from the hole-hole interaction which allows the radiative recombination of the momentum indirect electron-hole component of the trion (see text). **(c)** The second hole ends up in the Γ valley, almost degenerate with the B valence band in K valley.

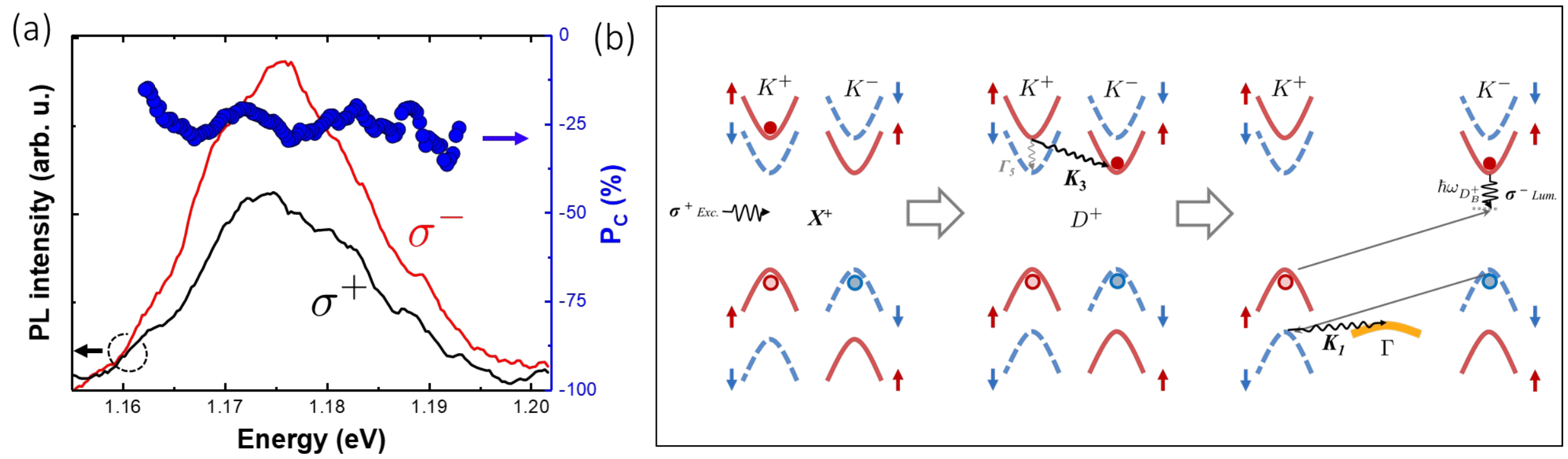

**Fig. 6 Circularly-polarized emission of the positively charged brightened dark trion. (a)** Right (σ+) and left (σ–) circularly-polarized luminescence of the $D_B^+$ line following σ+ polarized laser excitation; the corresponding circular polarization $P_c$ is also displayed (note its negative value). **(b)** Simplified scheme yielding a negative PL circular polarization of the $D_B^+$ line as a consequence of the brightening process (see text).